\documentclass[aps,prb,twocolumn,showpacs,a4paper]{revtex4}

\usepackage{amsfonts}
\usepackage{amsmath}
\usepackage{dcolumn}
\usepackage[english]{babel}
\usepackage{graphicx}
\usepackage[small,bf,hang,raggedright,FIGTOPCAP]{subfigure}
\usepackage{varioref}
\input epsf
\usepackage{epstopdf}

\begin{document}
\title{Elastic properties of hydrogenated graphene}
\author{Emiliano Cadelano$^1$, Pier Luca Palla$^2$, Stefano Giordano$^1$, Luciano Colombo$^1$}
\email[e-mail: ]{luciano.colombo@dsf.unica.it}
\affiliation{
$^1$ Dipartimento di Fisica, Universit\`a di Cagliari, Cittadella Universitaria, I-09042 Monserrato (Ca), Italy\\
$^2$ Dipartimento di Metodi e Modelli Matematici per le Scienze Applicate, Universit\`a di Padova, via Trieste 63, I-35121 Padova, Italy
}
\date{\today}

\begin{abstract}
There exist three conformers of hydrogenated graphene, referred to as chair-, boat-, or washboard-graphane. These systems have a perfect two-dimensional periodicity mapped onto the graphene scaffold, but they are characterized by a $sp^3$ orbital hybridization, have different crystal symmetry, and otherwise behave upon loading. By first principles calculations we determine their structural and phonon properties, as well as we establish their relative stability. Through continuum elasticity we define a simulation protocol addressed to measure by a computer experiment their linear and nonlinear elastic moduli and we actually compute them by first principles. We argue that all graphane conformers respond to any arbitrarily-oriented extention with a much smaller lateral contraction than the one calculated for graphene. Furthermore, we provide evidence that boat-graphane has a small and negative Poisson ratio along the armchair and zigzag principal directions of the carbon honeycomb lattice (axially auxetic elastic behavior). Moreover, we show that chair-graphane admits both softening and hardening hyperelasticity, depending on the direction of applied load.
\end{abstract}
\maketitle

\section{Introduction}
\begin{figure}[b]
\begin{center}
\includegraphics[width= 0.45\textwidth, angle=0]{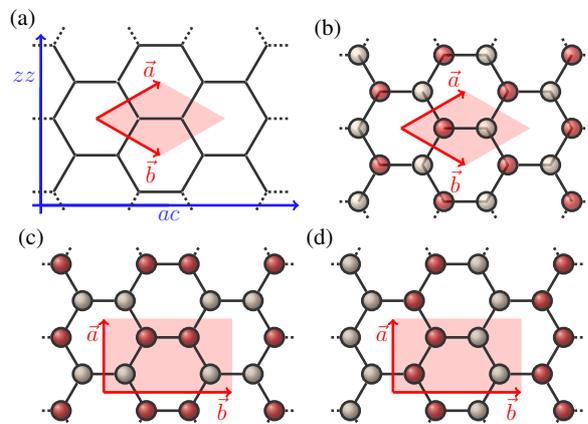}
\caption{\label{gra}(color online) Pictorial representations of the graphane conformers, obtained by different hydrogen decorations (the actual atomic positions are reported in Fig.~\ref{fig:structPerspective}). Top hydrogen atoms are indicated by red (dark) circles, while bottom ones by gray (light) circles. Shaded areas represent the unit cell and the corresponding lattice vectors are indicated by $\vec a$ and $\vec b$. Panel a: graphene scaffold (full lines) with zigzag (zz) and armchair (ac) directions. Panel b, c, and d:  chair-, boat-, and washboard-graphane, respectively.}
\end{center}
\end{figure}

The hydrogenated form of graphene is referred to as graphane. It is described as a two-dimensional, periodic, and covalently bonded hydrocarbon with a C:H ratio of 1. Hydrogen atoms decorate the carbon honeycomb lattice on both the top and bottom side (see Fig.~\ref{gra}).
Graphane was theoretically predicted by Sofo {\it et al.}, \cite{sofo} further investigated by Boukhvalov  {\it et al.} \cite{boukhvalov} and eventually grown by Elias {\it et al.} \cite{elias} The investigation of graphane properties was originally motivated by the search for novel materials with possibly large impact in nanotechnology.

The attractive feature of graphane is that by variously decorating the graphene atomic scaffold with hydrogen atoms (still preserving periodicity) it is in fact possible to generate a set of two dimensional materials with new physico-chemical properties. This is obviously due to change in the orbital hybridization which, because of hydrogenation, is now $sp^3$-like. For instance, it has been calculated \cite{sofo,boukhvalov} that graphane is an insulator, with an energy gap as large as $\sim3$ eV, while graphene is a highly conductive semi-metal. In case the hydrogenated sample is disordered, the resulting electronic and phonon properties are yet again different.\cite{elias}
Hydrogenation likely affects the elastic properties as well. Topsakal {\it et al.} \cite{topsakal} indeed calculated that the in-plane stiffness and Poisson ratio of graphane are smaller than those of graphene. In addition, the value of the yield strain is predicted to vary upon temperature and stoichiometry.

\begin{figure}[b]
\centering%
\subfigure[\label{chairPerspective}C-graphane]%
{\includegraphics[width= 0.2\textwidth]{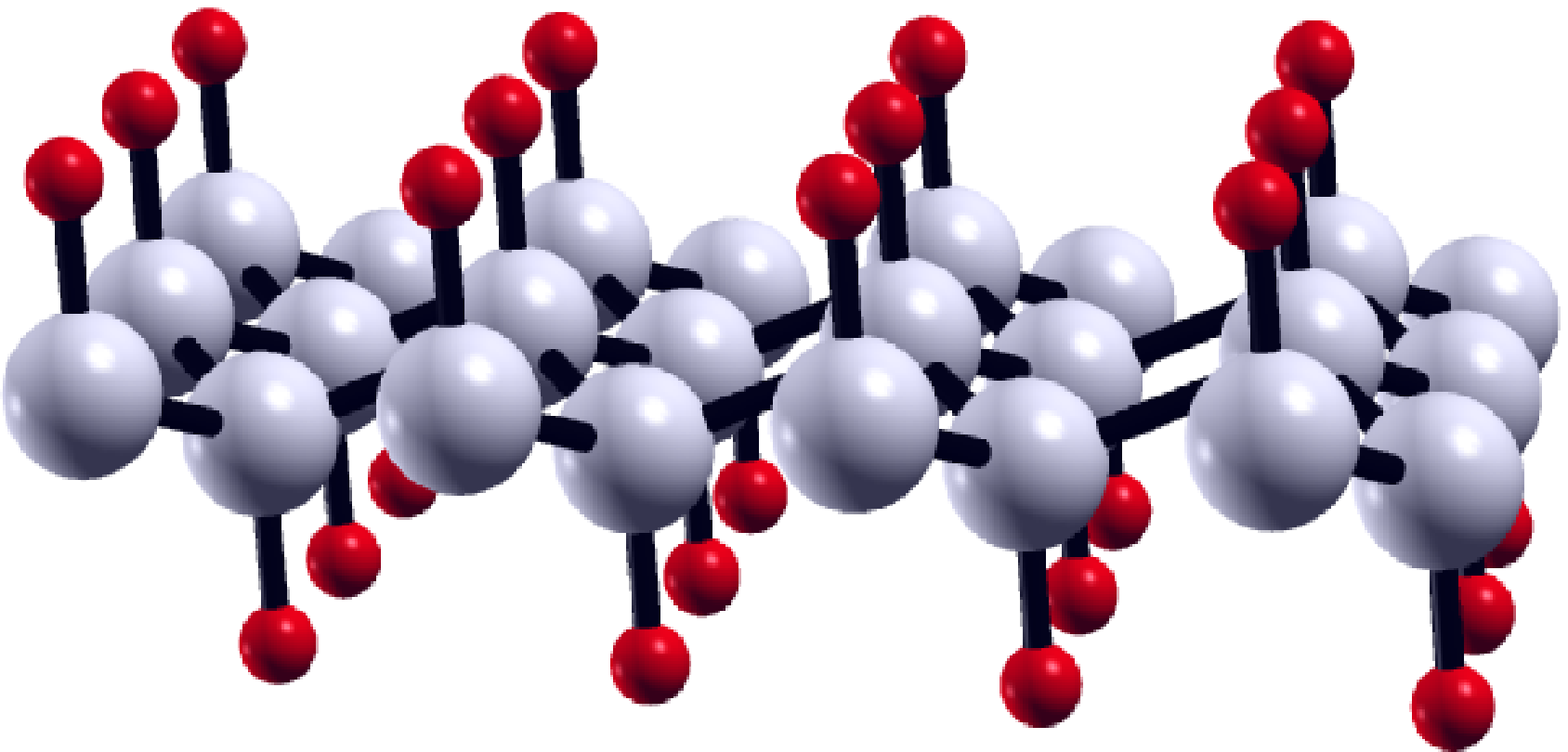}
\includegraphics[width= 0.25\textwidth]{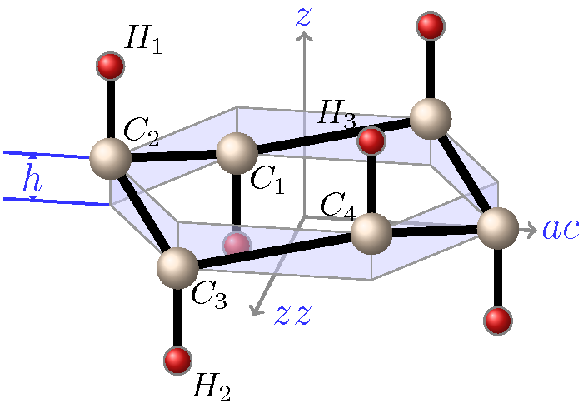}}\qquad
\subfigure[\label{boatPerspective}B-graphane]%
{\includegraphics[width= 0.2\textwidth]{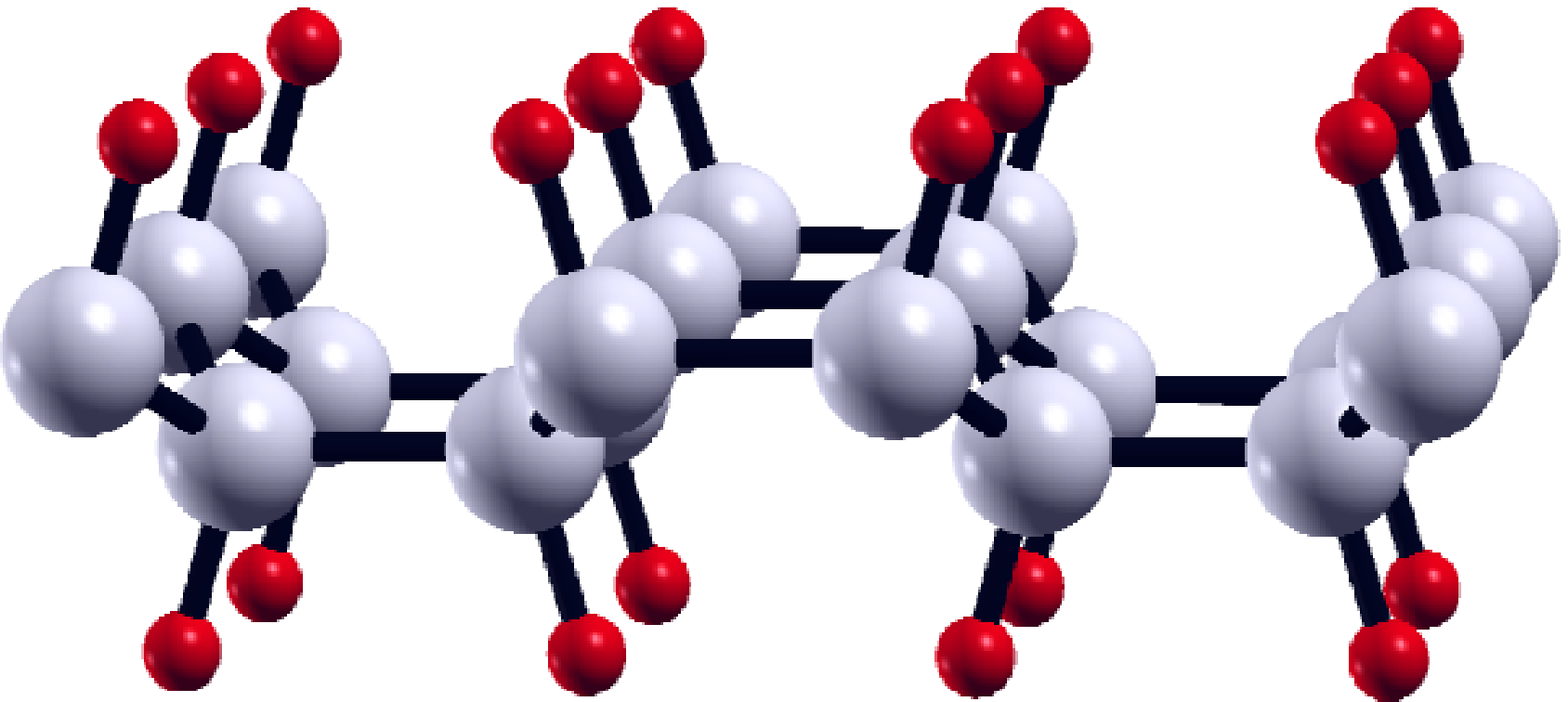}
\includegraphics[width= 0.25\textwidth]{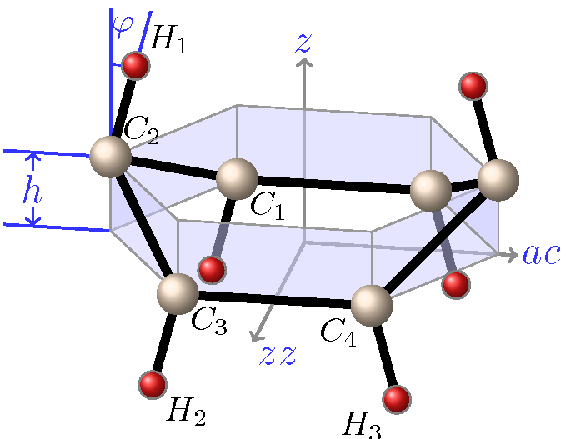}}\qquad%
\subfigure[\label{washboardPerspective}W-graphane]%
{\includegraphics[width= 0.2\textwidth]{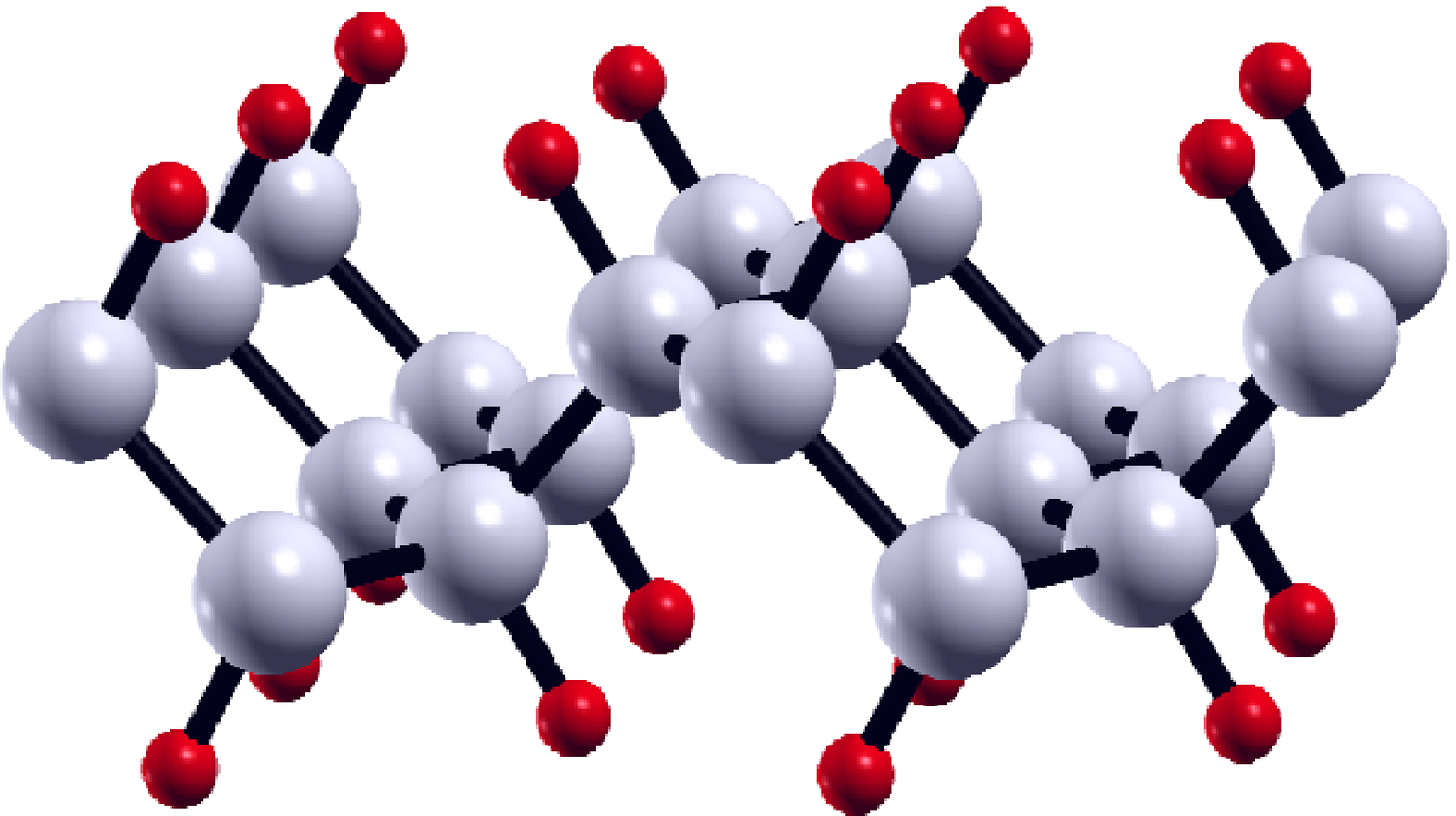}
\includegraphics[width= 0.25\textwidth]{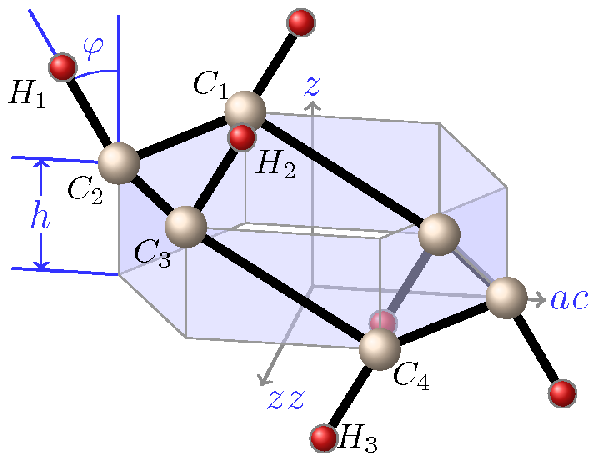}}\qquad%
\caption{(color online) Perspective representations of fully relaxed graphane conformers. Gray (light gray) and red (dark gray) spheres represent carbon and hydrogen atoms, respectively. Labels $C_n$ and $H_n$ (with $n=1, 2, 3$ and 4) provide the atom identifications used in Table \ref{table:structure}. Right panels show the orientation with respect to the armchair (ac) and zigzag (zz) direction, as well as the structural parameters $h$ and $\varphi$ reported in Table \ref{table:structure}. } 
\label{fig:structPerspective}
\end{figure}
As far as the mechanical properties of graphane are concerned, the $sp^2$-to-$sp^3$ change in orbital hybridization causes a major difference with respect to graphene. There in fact exist graphane conformers which are not isotropic, at variance with graphene which is so (in linear approximation \cite{cadelano}).
This feature stimulates an intriguing change of perspective, namely: hydrogenation could not only affect the actual value of some linear elastic moduli;\cite{topsakal} it could even dramatically change the overall mechanical behavior of the system by introducing an anisotropic dependence of its respose to an external load. This is in fact what we predict in this work by first principles total energy calculations, combined to continuum elasticity: we show that there is a graphane conformer (i.e., boat graphane as detailed below) showing a vanishingly small (possibly negative) Poisson ratio upon loading along given directions. In other words, we provide evidence that upon suitable hydrogenation a graphene sheet behaves as an axially auxetic material,\cite{evans} namely: it does not shrink, but actually slightly elongates perpendicularly to an applied traction force.
Nonlinear elastic features show an interesting anisotropic behavior as well. 

This paper is organized as follows. In Sec.~\ref{sec:methods}, the methods and the general computational setup adopted in our calculations are outlined. In Sec.~\ref{sec:structure} we provide a full structural characterization of three graphane conformers and we discuss their stability. In Sec.~\ref{sec:linear} and Sec.~\ref{sec:nonlinear} we describe their linear and nonlinear elastic properties, respectively, and we compute all the relevant elastic moduli.

\begin{table}[t]
\label{table:structure}
\caption{Space groups and structural parameters for each graphane conformers. The cell parameters $a$ and $b$ are defined in Fig.~\ref{gra} while the other quantities are reported in Fig.~\ref{fig:structPerspective}. 
Note that the B-graphane shows two types of C-C bonds while W-graphane exhibits a large buckling parameter, $h$.
}
\begin{tabular}{crrr}
\hline
\hline
\                      & \ \ C-graphane\ \  & \ \ B-graphane\ \  & \ \ W-graphane \\
\hline\noalign{\smallskip}
Space Group            &     P-3m1 (164)    & \ \  Pmmn  (59)    & \ \   Pmna (53)\\
\\
  $a$                  & 2.54    \AA        & 2.53	\AA      & 2.55 \AA 	\\
  $b$                  &   -\;\;\;\;        & 4.31	\AA      & 3.82 \AA 	\\
  $C_1-C_2$              & 1.54    \AA        & 1.54	\AA      & 1.54 \AA 	\\
  $C_3-C_4$              & 1.54    \AA        & 1.57 	\AA      & 1.54 \AA \\
  $C-H$                & 1.11    \AA        & 1.11      \AA      & 1.11	\AA \\
       $h$               & 0.46    \AA        & 0.65      \AA      & 1.14	\AA \\
\\
$\varphi$               & 0.0$^{\circ}$      & 16.7$^{\circ}$     & 30.1$^{\circ}$  \\
$\widehat{C_1 C_2 C_3}$   & 111.5$^{\circ}$    & 110.7$^{\circ}$	 & 111.2$^{\circ}$ \\
$\widehat{C_2 C_3 C_4}$   & 111.5$^{\circ}$   & 112.3$^{\circ}$	 & 112.3$^{\circ}$\\
$\widehat{H_1 C_2 C_3}$   & 107.4$^{\circ}$   & 107.2$^{\circ}$    & 106.5$^{\circ}$\\
\\
$\widehat{H_1 C_2 C_3 H_2}$& 180.0$^{\circ}$      & 180.0$^{\circ}$	 & 51.2$^{\circ}$ \\
$\widehat{H_2 C_3 C_4 H_3}$& 180.0$^{\circ}$      & 0.0$^{\circ}$      & 0.0$^{\circ}$\\
\hline
\hline
\end{tabular}
\end{table}
\section{Computational setup}
\label{sec:methods}

All calculations have been performed by Density Functional Theory (DFT) as implemented in the \textsc{Quantum ESPRESSO} package.\cite{quantum-espresso} The exchange correlation potential was evaluated through the generalized gradient approximation (GGA), using the Vanderbilt ultrasoft pseudopotential PW91.\cite{vanderbilt} A plane wave basis set with kinetic energy cutoff as high as 50 Ry was used and in most calculations the Brillouin zone (BZ) has been sampled by means of a (18x18x3) Monkhorst-Pack grid. The atomic positions of the investigated samples have been optimized by using the quasi-Newton algorithm and periodically-repeated simulation cells. Accordingly, the interactions between adjacent atomic sheets in the supercell geometry was hindered by a large spacing greater than 10 \AA.

The elastic moduli of the structures under consideration have been obtained from the energy-vs-strain curves, corresponding to suitable sets of deformations applied to a single unit cell sample. As discussed in more detail in Sec.~\ref{sec:linear} and Sec.~\ref{sec:nonlinear}, for any deformation the magnitude of the strain is represented by a single parameter $\zeta$. The curves have been carefully generated by increasing the magnitude of $\zeta$ in steps of  0.001 up to a maximum strain $|\zeta_{max}|=0.05$. All results have been confirmed by checking the stability of the estimated elastic moduli over several fitting ranges.
The reliability of the above computational set up is proved by the estimated values for the Young modulus and the Poisson ratio of graphene, respectively 344 Nm$^{-1}$ and 0.169, which are in excellent agreement with recent literature.\cite{kudin,gui,liu} Similarly, our results for the same moduli in C-graphane (respectively, 246 Nm$^{-1}$ and 0.08) agree very well with data reported in Ref. \onlinecite{topsakal}.

The stability of the three graphane conformers has been established by calculating the corresponding phonon dispersions. Phonon dispersions, have been obtained by means of Density-Functional Perturbation Theory (DFPT),\cite{baroni} based on the $(2n+1)$ theorem. In this case, during the self-consistent field calculation, the BZ has been sampled by a (16x16x3) Monkhorst-Pack grid. The accuracy of the phonon dispersion evaluations has been tested on a graphene sample (see below).

\section{Structure and stability of graphane conformers}
\label{sec:structure}
By hydrogenating a honeycomb graphene lattice, three ordered graphane structures can be generated, namely: the chair (C-graphane), boat (B-graphane)  and washboard (W-graphane) conformers \cite{sofo,washboard} shown in Fig.~ \ref{fig:structPerspective}.

Each conformer is characterized by a specific hydrogen sublattice and by a different buckling of the carbon sublattice. In particular: in C-graphane the hydrogen atoms alternate on both sides of the carbon sheet; in B-graphane pairs of H-atoms alternate along the armchair direction of the carbon sheet; finally, in W-graphane double rows of hydrogen atoms, aligned along the zigzag direction of the carbon sublattice, alternate on both sides of the carbon sheet.  A perspective view of
the conformers is shown in Fig.~ \ref{fig:structPerspective} and the corresponding structural data are given in Table \ref{table:structure}.
In C-graphane and W-graphane the calculated C-C bond length of 1.54 \AA ~is similar to the $sp^3$ bond length in diamond and much larger than in graphene. Moreover, we note that the B-graphane shows two types of C-C bonds, namely: 
those connecting two carbon atoms bonded to hydrogen atoms either lying on opposite sides (bond length 1.57 \AA) or lying on the same side of honeycomb scaffold (bond length 1.54 \AA).
Finally, the C-H bond length of 1.1 \AA ~ is similar in all conformers and it is typical of any hydrocarbon.

\begin{figure}[bp]
\centering%
\subfigure[\label{graphenePhon} graphene]%
{\includegraphics[width= 0.35\textwidth]{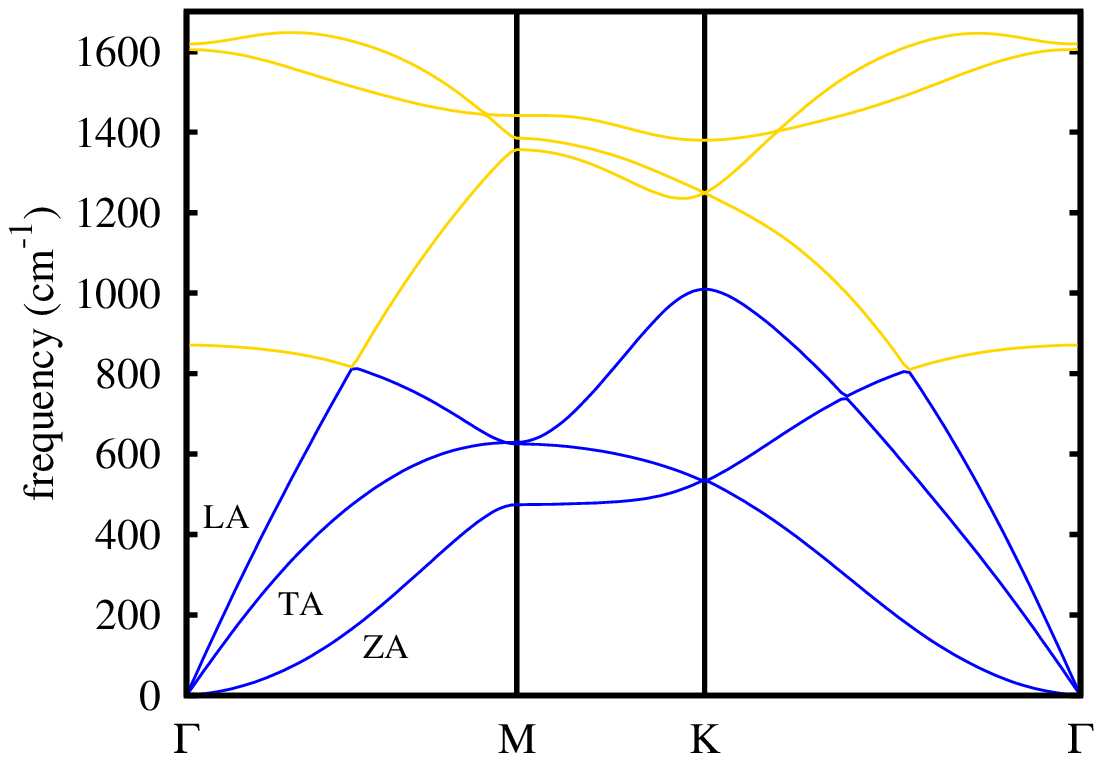}}\qquad
\subfigure[\label{chairPhon} C-graphane]%
{\includegraphics[width= 0.35\textwidth]{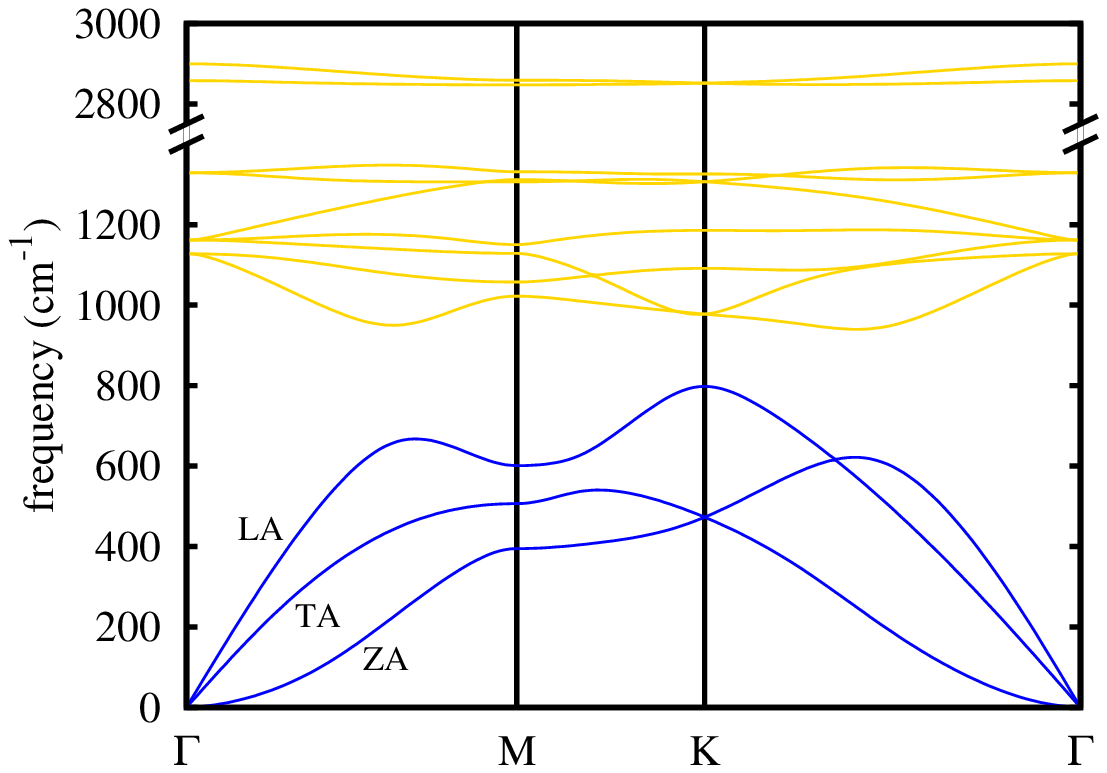}}\qquad%
\subfigure[\label{boatPhon} B-graphane]%
{\includegraphics[width= 0.35\textwidth]{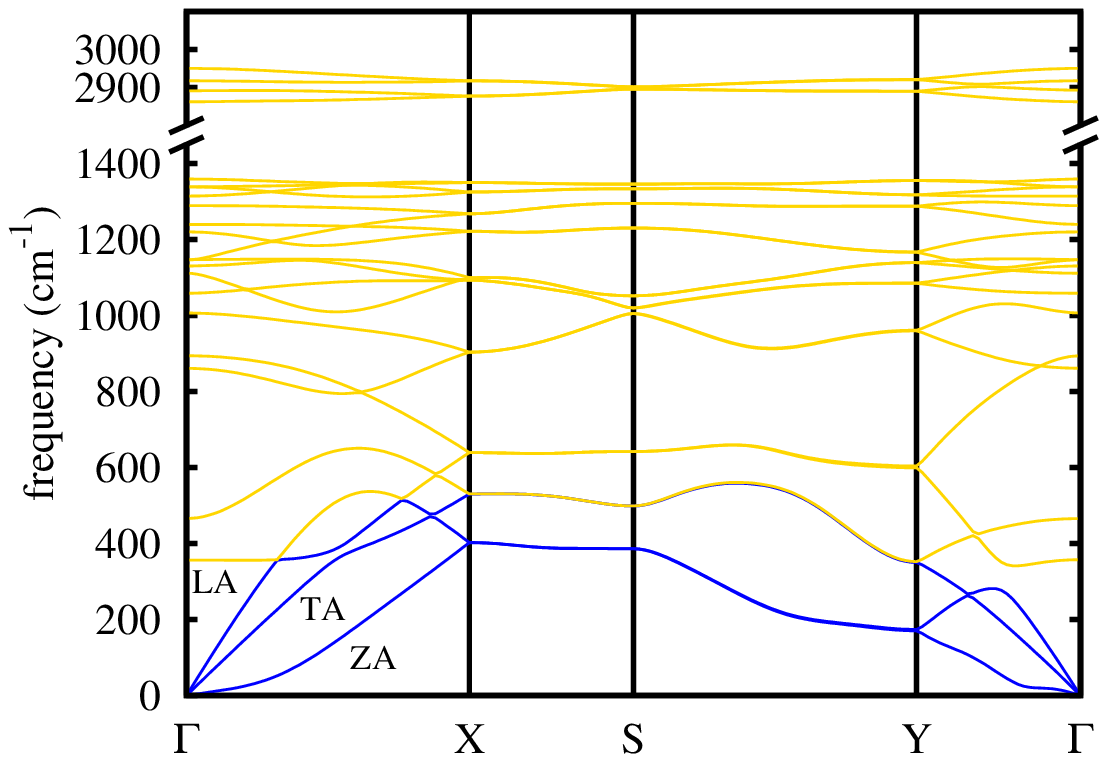}}\qquad%
\subfigure[\label{washboardPhon} W-graphane]%
{\includegraphics[width= 0.35\textwidth]{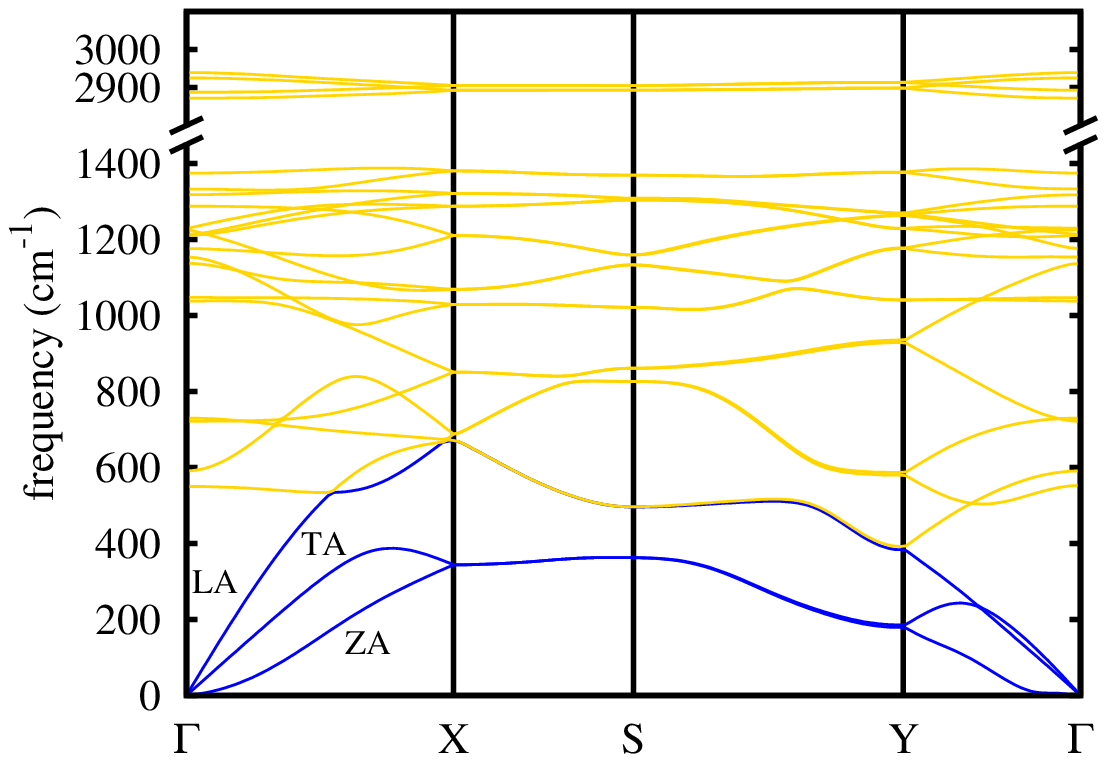}}\qquad%
\caption{\label{fig:phonon}(color online) Phonon dispersion relations of graphene (panel a), C- (panel b), B- (panel c) and  W-  (panel d) graphane. Acoustic  and optical modes correspond, respectively, to blue (dark gray) and yellow (light gray) dispersions. Longitudinal and transverse acoustic branches are indicated as LA and TA, respectively. The acoustic branch with displacement patterns along the $z$-direction of Fig.~ \ref{fig:structPerspective} is marked as ZA.}
\label{fig:structPhon}
\end{figure}

The stability of the three graphane conformers has been established by calculating the phonon dispersion curves reported in Fig.~ \ref{fig:phonon}. Graphene phonon spectrum is reported as well for comparison.
No soft modes (with negative frequency) corresponding to possible instabilities were found along any high-symmetry direction of the Brillouin zone. Furthermore, as expected,\cite{dresselhaus} the zone-center longitudinal (LA) and transverse (TA) acoustic branches show a linear dependence upon the wavevector, while the acoustic mode ZA (with displacement patterns along the 
$z$-direction shown in Fig.~ \ref{fig:structPerspective}) shows a quadratic dependence.
We observe that in C-graphane, as well as in graphene, the speed of sound (i.e. the slope of the acoustic branches at $\Gamma$-point) is the same along the $\Gamma-M$ and $\Gamma-K$ directions. On the other hand, the B- and W-graphane conformers are characterized by different sound velocities along the $\Gamma-X$ and $\Gamma-Y$ directions. This is the fingerprint of an unlike elastic behavior: as extensively discussed in Sec.~\ref{sec:linear}, C-graphane is elastically isotropic while neither B- nor W-graphane are so.
 
Finally, according to the present first principles total energy calculations we identified C-graphane as the most energetically favorable conformer. W- and B-graphane have higher ground-state energy  of 0.05 and 0.10 eV (per C-H unit), respectively. These small differences in energy demonstrate that all three conformers are thermodynamically accessible, as indeed experimentally guessed.\cite{washboard}

\section{Linear elasticity}
\label{sec:linear}
While C-graphane has trigonal symmetry (and, therefore, is elastically isotropic as hexagonal graphene), the remaining B- and W-conformers show an orthorhombic symmetry, which causes an anisotropic linear elastic behavior. Accordingly, the elastic energy density (per unit of area) accumulated upon strain can be expressed as \cite{huntington}
\begin{eqnarray}
U_{trigo}= \frac{1}{2}C_{11}\left( \epsilon_{xx}^{2}+\epsilon_{yy}^{2}+2\epsilon_{xy}^{2}\right) +C_{12}\left(  \epsilon_{xx}\epsilon_{yy}-\epsilon_{xy}^{2}\right) 
\label{trigo}
\end{eqnarray}
for the isotropic structures and as
\begin{eqnarray}
U_{ortho}= \frac{1}{2}C_{11}\epsilon_{xx}^{2}+\frac{1}{2}C_{22}\epsilon_{yy}^{2}+C_{12} \epsilon_{xx}\epsilon_{yy}+2C_{44}\epsilon_{xy}^{2}
\label{orto}
\end{eqnarray}
for the anisotropic ones.
In Eqs.(\ref{trigo}) and (\ref{orto}) we have explicitly made use of the elastic linear constants  $ C_{11}$, $C_{22} $, $ C_{12}$ and $C_{44} $. Furthermore, the infinitesimal strain tensor $ \hat \epsilon =\frac{1}{2}(\vec{\nabla}\vec{u}+\vec{\nabla}\vec{u}^{\rm T})$ is represented by a symmetric matrix with elements $ \epsilon_{xx}=\frac{\partial u_x}{\partial x} $, $ \epsilon_{yy}=\frac{\partial u_y}{\partial y} $ and $ \epsilon_{xy}=\frac{1}{2}\left(\frac{\partial u_x}{\partial y}+ \frac{\partial u_y}{\partial x} \right) $, where the functions $ u_{x}(x,y) $ and $ u_{y}(x,y) $ correspond to the planar displacement $ \vec{u}=(u_x,u_y) $.
It is important to remark that $U_{trigo}$ can be obtained from the $U_{ortho}$ by simply imposing the isotropy condition $ C_{11}=C_{22} $ and the Cauchy relation $ 2C_{44}=C_{11}-C_{12} $, holding for both the hexagonal and trigonal symmetry. We will take profit of this by focussing just on the elastic behavior of a system described by Eq.(\ref{orto}); when needed, the general results so obtained will be applied to the isotropic structures by fully exploiting the above conditions. The constitutive in-plane stress-strain equations are straightforwardly derived from Eq.(\ref{orto}) through $\hat{T}= \partial U/ \partial \hat{\epsilon}$, where $ \hat{T} $ is the Cauchy stress tensor \cite{landau}. They are: $ T_{xx}=C_{11} \epsilon_{xx} +C_{12} \epsilon_{yy}$, $  T_{yy}=C_{22} \epsilon_{yy} +C_{12} \epsilon_{xx} $ and $ T_{xy}=2C_{44} \epsilon_{xy} $.

We now suppose to apply an axial tension $ \sigma $ to any  two dimensional hydrocarbon shown in Fig.~ \ref{fig:structPerspective} along the arbitrary direction $ \vec{n}=\cos\theta \vec{e}_{x}+\sin\theta \vec{e}_{y} $, where $ \vec{e}_{x} $ and $ \vec{e}_{y} $ are, respectively, the unit vectors along the zigzag and the armchair directions of the underlying honeycomb lattice. In this notation, therefore, $\theta$ is the angle between $\vec{n}$ and the zigzag direction. 
Under this assumption we get $\hat{T}=\sigma  \vec{n}\otimes\vec{n}$, where the in-plane stress components are defined, respectively, as $ T_{xx}=\sigma \cos^{2}\theta $, $T_{xy}=\sigma\cos\theta\sin\theta$, and $ T_{yy}=\sigma \sin^{2}\theta $. By inverting the constitutive equation we find the corresponding strain tensor $ \hat{\epsilon} $. In particular, we easily get its longitudinal component $\epsilon_{l}= \vec{n}\cdot \hat{\epsilon}\ \vec{n} $ along the direction $ \vec{n} $
\begin{equation}
\epsilon_{l}= \sigma \left[ \frac{C_{11}}{\Delta}s^4+\frac{C_{22}}{\Delta}c^4 +\left( \frac{1}{C_{44}}-2 \frac{C_{12}}{\Delta}\right)c^2 s^2 \right]
\end{equation}
as well as its transverse component $\epsilon_{t}= \vec{t}\cdot \hat{\epsilon}\ \vec{t} $ along the direction $ \vec{t}=-\sin\theta \vec{e}_{x}+\cos\theta \vec{e}_{y} $ 
(with $ \vec{t}\cdot\vec{n}=0 $)
\begin{equation}
\epsilon_{t}=\sigma \left[ \left( \frac{C_{11}+C_{22}}{\Delta}-\frac{1}{C_{44}}\right) c^2 s^2-\frac{C_{12}}{\Delta}\left(c^4 + s^4\right) \right]
\end{equation}
where $ \Delta=C_{11}C_{22}-C_{12}^2$, $ c=\cos \theta$, and 
$ s=\sin \theta $.
By means of Eqs.(3) and (4) we obtain, respectively, the $\vec{n}$-dependent Young modulus 
$E_{\vec{n}}=\sigma/ \epsilon_l $ (i.e. the ratio between the applied traction and the longitudinal extension) as
\begin{equation}
\label{EEE}
E_{\vec{n}}= \frac{\Delta}{C_{11}s^4+C_{22}c^4 +\left( \frac{\Delta}{C_{44}}-2 {C_{12}}\right)c^2 s^2}
\end{equation}
and the $\vec{n}$-dependent Poisson ratio $ \nu_{\vec{n}}=-\epsilon_t / \epsilon_l $ (i.e. the ratio between the lateral contraction and the longitudinal extension) as
\begin{equation}
\label{nunu}
\nu_{\vec{n}}=-\frac{\left( {C_{11}+C_{22}}-\frac{\Delta}{C_{44}}\right) c^2 s^2-{C_{12}}\left(c^4 + s^4\right) }{C_{11}s^4+C_{22}c^4 +\left( \frac{\Delta}{C_{44}}-2 {C_{12}}\right)c^2 s^2} 
\end{equation}
Eqs.(\ref{EEE}) and (\ref{nunu}) are central to our investigation. 

First of all, we remark that they allow for the full linear elastic characterization of both the anisotropic graphane conformers and the trigonal one (as well as graphene), provided that in the latter case the isotropy and Cauchy conditions are duly exploited. In this case we in fact obtain the Young modulus 
$E=(C_{11}^2-C_{12}^2)/C_{11} $ and the Poisson ratio $ \nu=C_{12}/C_{11} $, which are independent of the angle $ \theta $, confirming the planar isotropy.
\begin{table*}[t]
\caption{\label{strain-tensor} Deformations and corresponding strain tensors applied to compute the elastic constants $C_{ij}$ of graphane. The relation between such constants and the fitting term $U^{(2)}$ of Eq.(\ref{fit}) is reported as well. Deformations (i)-(ii) are applied to the C-conformer, while the full set (i)-(iv) of deformations is applied to the B- and W-conformers. $\zeta$ is the scalar strain parameter.}
\begin{tabular}{lrcccc}
\hline
\hline\noalign{\smallskip}
\ & \ \ \ & strain tensor & $U^{(2)}$ & \ \ \ & $U^{(2)}$ \\
\ & \ & \ \ \ &  isotropic structures & \ \ \ & anisotropic structures \\
\hline\noalign{\smallskip}
\ (i) zigzag axial deformation & \ \ \ & $\bigl(\begin{smallmatrix} \zeta & 0  \\0 & 0 \end{smallmatrix} \bigr)$ & $C_{11}$ & \ \ \ & $C_{11}$  \vspace{.5ex}\\
(ii) hydrostatic planar deformation & \ \ \ & $\bigl(\begin{smallmatrix} \zeta & 0  \\0 & \zeta \end{smallmatrix} \bigr)$ &$2(C_{11}+C_{12})$ & \ \ \ & $C_{11}+C_{22}+2C_{12}$ \vspace{.5ex}\\
(iii) armchair axial deformation & \ \ \ & $\bigl(\begin{smallmatrix} 0 & 0  \\0 & \zeta \end{smallmatrix} \bigr)$ & \ & \ \ \ & $C_{22}$ \vspace{.5ex}\\
(iv) shear deformation & \ \ \ & $\bigl(\begin{smallmatrix} 0 & \zeta  \\ \zeta & 0 \end{smallmatrix} \bigr)$ & \ & \ \ \ & $4C_{44}$ \vspace{.5ex}\\
\hline
\hline
\end{tabular}
\end{table*}

More importantly, however, Eqs.(\ref{EEE}) and (\ref{nunu}) imply that $E_{\vec{n}}$ and 
$\nu_{\vec{n}}$ can be directly obtained by the linear elastic constants $C_{ij}$, in turn computed through energy-vs-strain curves corresponding to suitable homogeneous in-plane deformations. This implies that there is no actual need to mimic by a computer simulation a traction experiment along the arbitrary direction identified by $ \vec{n} $ or $ \theta$, indeed a technically complicated issue to accomplish. Rather, for the isotropic case (graphene and C-graphane) only two in-plane deformations should be applied in order to obtain all the relevant elastic constants, namely: (i) an axial deformation along the zigzag direction; and (ii) an hydrostatic planar deformation. For the anisotropic case (B- and W-graphane)  two more in-plane deformations must be applied: (iii) an axial deformation along the armchair direction; and (iv) a shear deformation. The strain tensors corresponding to deformations (i)-(iv) depend by a unique scalar strain parameter $ \zeta $ as shown in Table \ref{strain-tensor}. For all imposed deformations the elastic energy of strained structures can be written in terms of 
$\zeta$ as
\begin{equation}
 \label{fit}	
U(\zeta)=U_0+\frac{1}{2}U^{(2)}\zeta^2+O(\zeta^3)
\end{equation}
where $U_0$ is the energy of the unstrained configuration. Since the expansion coefficient $U^{(2)}$ is related to the elastic moduli as summarized in Table \ref{strain-tensor}, a straightforward fit of 
Eq.(\ref{fit}) has provided the full set of linear moduli for all structures.

\begin{table}[b]
\caption{\label{tablefin} Graphene and graphane independent elastic constants (units of  Nm$^{-1}$). For graphene and C-graphane $ C_{11}=C_{22} $ and $ 2C_{44}=C_{11}-C_{12} $.}
\begin{tabular}{crrrr}
\hline
\hline
\                & \ \ graphene\ \  & \ \ C-graphane\ \  & \ \ B-graphane\ \  & \ \ W-graphane \\
\hline\noalign{\smallskip}
$C_{11}$  & 354          & 248              &  258              & 280 \\
$C_{22}$  &         &            &  225              & 121 \\
$C_{12}$  &   60          &   20              & -1.7               &  14  \\
$C_{44}$  &         &               &   93               &   81 \\
\hline
\hline
\end{tabular}
\end{table}
\begin{figure}[b]
\begin{center}
\includegraphics[width= 0.4\textwidth, angle=0]{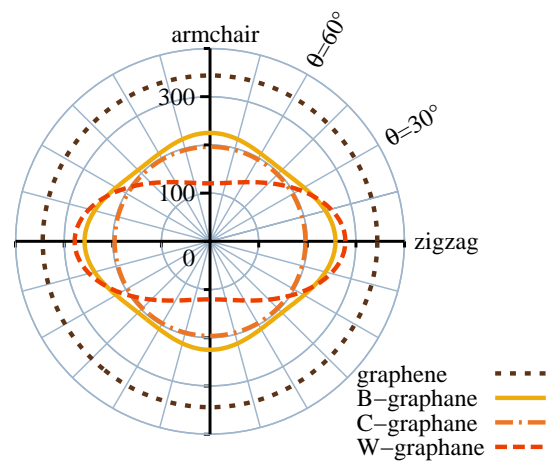}
\caption{\label{young} (color online) Polar diagram for the Young modulus $E$ of graphene and graphane conformers. The angle $\theta$ identifies the extension direction with respect to the zigzag one. Isotropic (anisotropic) behavior is associated to a circular (non circular) shape of the $E_{\vec{n}}$ plot.}
\end{center}
\end{figure}

The synopsis of the calculated elastic constants for all graphane conformers, as well as graphene, is reported in Table \ref{tablefin}, from which three qualitative information can be extracted. First, we observe that the difference between $C_{11}$ and $C_{22} $ is much smaller for the B-conformer than for W-graphane; therefore, this latter is by far the most elastically anisotropic conformer. Then, the value of $C_{44}$, measuring the resistance to a shear deformation, decreases monotonically from graphene to W-graphane. Finally, we remark that the value of $C_{12}$ (or, similarly, of the Poisson ratio) is much smaller in any graphane structure than in pristine graphene. The change in hybridization has therefore largely reduced the property of lateral contraction upon extension. Interestingly enough, the B-conformer is characterized by a negative $C_{12}$ value, something unexpected and worthy of further investigation, as reported below.
 
\begin{figure}[t]
\begin{center}
\includegraphics[width= 0.4\textwidth, angle=0]{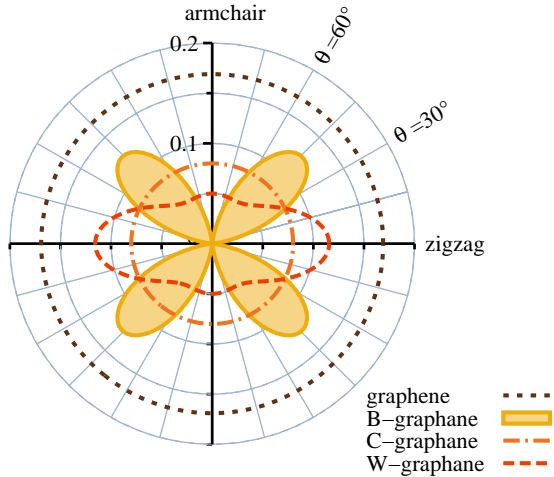}
\caption{\label{poisson} (color online) Polar diagram for the Poisson ratio $\nu$ of graphene and graphane conformers. The angle $\theta$ identifies the extension direction with respect to the zigzag one. Isotropic (anisotropic) behavior is associated to a circular (non circular) shape of the $\nu_{\vec{n}}$ plot. The special case of B-graphane is enlighten by shading (see text).}
\end{center}
\end{figure}

Through Eqs.(\ref{EEE}) and (\ref{nunu}) and by using the elastic constants reported in Table \ref{tablefin}, we can quantify the $\vec{n}$-dependence of $E$ and $\nu$ for the anisotropic structures by using polar coordinates, as illustrated in Fig.~\ref{young} and Fig.~\ref{poisson}, respectively. In such a representation, a fully isotropic elastic behavior is represented by a perfectly circular shape of the $E_{\vec{n}}$ and $\nu_{\vec{n}}$ plots. This is indeed the case, as expected, of graphene and C-graphane. On the other hand,
Fig.~\ref{young} confirms that W-graphane is much more anisotropic than the B-conformer. Furthermore, as anticipated, Fig.~\ref{poisson} provides evidence that the Poisson ratio in any graphane conformer is much smaller than in pristine graphene, since the corresponding $\nu_{\vec{n}}$ polar plots are contained within the graphene circle.

\begin{figure}[b]
\begin{center}
\includegraphics[width= 0.4\textwidth, angle=0]{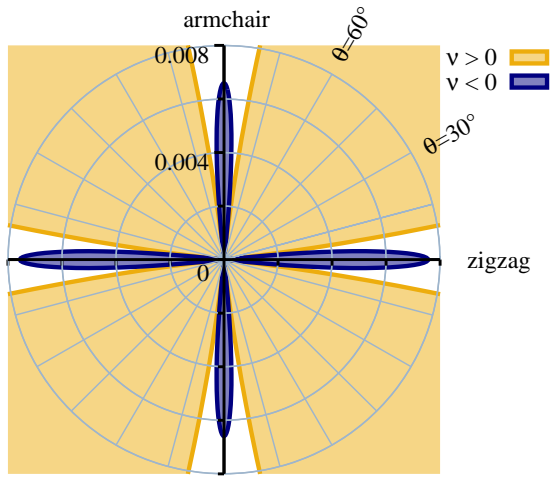}
\caption{\label{poissonzoom} (color online) The same as Fig.~ \ref{poisson} zoomed in the region nearby the origin. Positive and negative Poisson ratio values are differently shaded as indicated.}
\end{center}
\end{figure}

An intriguing unconventional behavior is observed in Fig.~\ref{poisson} for B-graphane, namely: for extensions along to the zigzag and armchair directions, the corresponding Poisson value is vanishingly small. This feature appears as a flower petal structure of the $\nu_{\vec{n}}$ plot for such a system.
By considering Fig.~\ref{poissonzoom}, where a zoom of the previous plot nearby the origin has been reported, we can actually learn more information. It is evident that four small lobes appear along the zigzag and armchair directions (i.e. along the principal axis of the orthorhombic symmetry), corresponding to a Poisson ratio varying in the range 
$ -0.0075<\nu<-0.0065 $. The limiting values are computed for extensions along the zigzag and armchair directions, respectively. It is truly remarkable that $\nu$ could be negative in B-graphane.
While a negative Poisson ratio value is allowed by thermo-elasticity, this peculiar situation is only observed in special systems (i.e. foams, molecular networks or tailored engineering structures) or just rarely in ordinary bulk materials (i.e. SiO$_2$, cubic metals, or polymer networks) \cite{advmat}. 

\section{Nonlinear elasticity}
\label{sec:nonlinear}
In this Section we generalize the previous analysis in order to draw a comparison between the nonlinear elastic behavior of graphene\cite{cadelano} and the three conformers of graphane. 
The nonlinear strain energy function $U_{hex}$ for an hexagonal two dimensional lattice is\cite{cadelano}
\begin{eqnarray}
\label{NLhex}
\nonumber
U_{hex}&=&  \frac{1}{2}C_{11}\left( \epsilon_{xx}^{2}+\epsilon_{yy}^{2}+2\epsilon_{xy}^{2}\right) +C_{12}\left(  \epsilon_{xx}\epsilon_{yy}-\epsilon_{xy}^{2}\right)\\ \nonumber
&+&\frac{1}{6} C_{111}\epsilon_{xx}^{3}+\frac{1}{6} C_{222}\epsilon_{yy}^{3} +\frac{1}{2}C_{112} \epsilon_{xx}^{2} \epsilon_{yy}\\
\nonumber
&+&\frac{1}{2}(C_{111}-C_{222}+C_{112})\epsilon_{xx} \epsilon_{yy}^{2}\\&+&\frac{1}{2}(3C_{222}-2C_{111}-C_{112})\epsilon_{xx} \epsilon_{xy}^{2}\nonumber \\
\label{completehex}
&+&\frac{1}{2}(2C_{111}-C_{222}-C_{112})\epsilon_{yy} \epsilon_{xy}^{2}
\end{eqnarray}
where all the nonlinear features are described by the three independent moduli $ C_{111}$, $C_{222}$ and $C_{112}$.\\
\begin{table}[t]
\caption{\label{graphanechair} Strain fields applied to compute the linear ($C_{ij}$) and nonlinear ($C_{ijk}$) elastic constants of the C-graphane.
The relation between such constants and the fitting terms $U^{(2)}$ and $U^{(3)}$ of Eq.(\ref{fit3}) is reported as well.
}
\begin{tabular}{lll}
\hline
\hline\noalign{\smallskip}
Strain&$U^{(2)}$&$U^{(3)}$\\
tensor&&\\
\hline\noalign{\smallskip}
\vspace{.5ex}
$\bigl(\begin{smallmatrix} \zeta & 0  \\0 & 0 \end{smallmatrix} \bigr)$ & $ C_{11} $&\: $ C_{111} $\\
\vspace{.5ex}
$\bigl(\begin{smallmatrix} \zeta & 0  \\0 & \zeta \end{smallmatrix} \bigr)$ & $ 2\left( C_{11} +C_{12}\right) $ &\: $ 2C_{111}+6C_{112} $\\
\vspace{.5ex}
$\bigl(\begin{smallmatrix} 0 & \zeta  \\ \zeta & 0 \end{smallmatrix} \bigr)$ & $ 2\left( C_{11} -C_{12} \right) $&\: $ 8C_{444} $\\
\vspace{.5ex}
$\bigl(\begin{smallmatrix} \zeta & \zeta  \\ \zeta & 0 \end{smallmatrix} \bigr)$ & $ 3C_{11}-2C_{12} $&\: $ C_{111}+12C_{144}+6C_{114}+8C_{444} $\\
\vspace{.5ex}
$\bigl(\begin{smallmatrix} 0 & \zeta  \\ \zeta & -\zeta \end{smallmatrix} \bigr)$ & $ 3C_{11}-2C_{12}  $&\: $ -C_{111}-12C_{144}+6C_{114}+8C_{444} $\\
\vspace{.5ex}
$\bigl(\begin{smallmatrix} \zeta & \zeta  \\ \zeta & -\zeta\end{smallmatrix} \bigr)$ & $ 4\left( C_{11}-C_{12}\right)  $ &\: $ 12C_{114}-12C_{124}+8C_{444}  $
\vspace{.5ex}\\
\hline
\hline
\end{tabular}
\end{table}
\begin{table}[b]
\caption{\label{graphaneboat} Strain fields applied to compute the linear ($C_{ij}$) and nonlinear ($C_{ijk}$) elastic constants of the B- and W-graphane.
The relation between such constants and the fitting terms $U^{(2)}$ and $U^{(3)}$ of Eq.(\ref{fit3}) is reported as well.
}
\begin{tabular}{lll}
\hline
\hline\noalign{\smallskip}
Strain &$U^{(2)}$&$U^{(3)}$\\
tensor&&\\
\hline\noalign{\smallskip}
$\bigl(\begin{smallmatrix} \zeta & 0  \\0 & 0 \end{smallmatrix} \bigr)$ & $ C_{11} $&\: $ C_{111} $\vspace{.5ex}\\
$\bigl(\begin{smallmatrix} 0 & 0  \\0 & \zeta \end{smallmatrix} \bigr)$ & $ C_{22} $&\: $ C_{222} $\vspace{.5ex}\\
$\bigl(\begin{smallmatrix} \zeta & 0  \\0 & \zeta \end{smallmatrix} \bigr)$ & $ C_{11} +C_{22}+2C_{12}$&\: $ C_{111}+C_{222}+3C_{112}+3C_{122} $\vspace{.5ex}\\
$\bigl(\begin{smallmatrix} 0 & \zeta  \\ \zeta & 0 \end{smallmatrix} \bigr)$ & $ 4C_{44} $&\: $ 0 $\vspace{.5ex}\\
$\bigl(\begin{smallmatrix} \zeta & \zeta  \\ \zeta & 0 \end{smallmatrix} \bigr)$ & $ C_{11}+4C_{44} $&\: $ C_{111}+12C_{144} $\vspace{.5ex}\\
$\bigl(\begin{smallmatrix} 0 & \zeta  \\ \zeta & \zeta \end{smallmatrix} \bigr)$ & $ C_{22}+4C_{44} $&\: $ C_{222}+12C_{244} $\vspace{.5ex}\\
$\bigl(\begin{smallmatrix} \zeta & 0  \\0 & -\zeta\end{smallmatrix} \bigr)$ & $ C_{11}+ C_{22}-2C_{12} $ &\: $ C_{111}-C_{222}-3C_{112}+3C_{122}  $\vspace{.5ex}\\
\hline
\hline
\end{tabular}
\end{table}
Similarly, the strain energy function  $U_{trigo}$ for C-graphane depending on the linear ($ C_{11} $ and $ C_{12} $) and nonlinear ($ C_{111}, C_{112}, C_{144}, C_{114},  C_{124} $ and $ C_{444} $) elastic constants is found to be
\begin{eqnarray}
\label{NLtrigo}
\nonumber
U_{trigo}&=& \frac{1}{2}C_{11}\left( \epsilon_{xx}^{2}+\epsilon_{yy}^{2}+2\epsilon_{xy}^{2}\right) +C_{12}\left(  \epsilon_{xx}\epsilon_{yy}-\epsilon_{xy}^{2}\right) \\ \nonumber
&+&\frac{1}{6} C_{111}\left( \epsilon_{xx}^{3}+\epsilon_{yy}^{3}\right) +\frac{1}{2}C_{112} \left( \epsilon_{xx}^{2} \epsilon_{yy}+\epsilon_{xx} \epsilon_{yy}^{2}\right) \\ \nonumber &+&2C_{144}\left( \epsilon_{xx} \epsilon_{xy}^{2}+\epsilon_{yy} \epsilon_{xy}^{2}\right) +C_{114}\left( \epsilon_{xx}^{2} \epsilon_{xy}+\epsilon_{yy}^{2} \epsilon_{xy}\right) \nonumber \\ &+& 2C_{124} \epsilon_{xx} \epsilon_{xy}\epsilon_{yy}+\frac{4}{3}C_{444}\epsilon_{xy}^{3}
\label{completetrigo}
\end{eqnarray}
For such a trigonal symmetry we have $ C_{111}=C_{222} $, $ C_{112}=C_{122}  $ and $ C_{144}=C_{244}  $. Nevertheless, it is important to underline that the overall nonlinear elastic response is truly anisotropic since not all the relevant isotropic conditions are fulfilled. 
\begin{table}[b]
\caption{ \label{table:nonlin} Graphene and graphane independent nonlinear elastic constants (units of  
Nm$^{-1}$). }
\begin{tabular}{ccccc}
\hline
\hline
\                & \ \ graphene\ \  & \ \ C-graphane\ \  & \ \ B-graphane\ \  
& \ \ W-graphane \\
\hline
$C_{111}$  &  -1910 $\pm$ 11        &  -1385$\pm$18          
&     -1609$\pm$31     &  -1756$\pm$33\\
$C_{222}$  &  -1764 $\pm$ 3         &                         
&     -1827$\pm$7       &  -487$\pm$85\\
$C_{112}$  &  -341  $\pm$ 35       &   -195$\pm$41          
&     -20$\pm$14        &    -75$\pm$54\\
$C_{122}$  &           &                         &     -55$\pm$22        
&    -296$\pm$36\\
$C_{124}$  &           &  -411$\pm$17           &                       
&   \\
$C_{114}$  &           &   530$\pm$12           &                        
&    \\
$C_{144}$  &           &   568$\pm$7            &     -161$\pm$4        
&    -143$\pm$17\\
$C_{244}$  &           &                              &     -159$\pm$3        
&    -287$\pm$10\\
$C_{444}$  &           &        0.0$\pm 10^{-5}$                       &            
&    \\
\hline
\hline
\end{tabular}
\end{table}

Finally, the strain energy function $U_{ortho}$ for the B- and W-graphane, expressed through the linear ($ C_{11}, C_{22}, C_{12}$ and $ C_{44} $) and nonlinear ($ C_{111}, C_{222}, C_{112}, C_{122}, C_{144}  $ and $ C_{244} $) elastic constants, is given by\\
\begin{eqnarray}
\label{NLortho}
\nonumber
U_{ortho}&=& \frac{1}{2}C_{11}\epsilon_{xx}^{2}+\frac{1}{2}C_{22}\epsilon_{yy}^{2}+2C_{44}\epsilon_{xy}^{2}
+C_{12} \epsilon_{xx}\epsilon_{yy}\\ &+&\frac{1}{6} C_{111}\epsilon_{xx}^{3}+\frac{1}{6} C_{222}\epsilon_{yy}^{3}+\frac{1}{2}C_{112} \epsilon_{xx}^{2} \epsilon_{yy} \nonumber \\&+&\frac{1}{2}C_{122}\epsilon_{xx} \epsilon_{yy}^{2}+2C_{144}\epsilon_{xx} \epsilon_{xy}^{2}+2C_{244}\epsilon_{yy} \epsilon_{xy}^{2}\,\,\,\,\,\,\,\,\,\,\,\,
\label{completeortho}
\end{eqnarray}
Eqs.(\ref{NLhex}), (\ref{NLtrigo}) and (\ref{NLortho}) can be obtained by using the standard tables of the tensor symmetries, found in many crystallography textbooks (see for instance Ref. \onlinecite{huntington}).

As above described, in any symmetry the strain energy function depends on the third-order elastic constants
(as well as the linear ones). Once again, they can be computed
through energy-vs-strain curves corresponding to suitable
homogeneous in-plane deformations. For each deformation the elastic
energy of strained graphene or graphane can be written in terms of just
the single deformation parameter $ \zeta$	
\begin{equation}
 \label{fit3}	
U(\zeta)=U_0+\frac{1}{2}U^{(2)}\zeta^2+\frac{1}{6}U^{(3)}\zeta^3+ O(\zeta^4)
\end{equation}
Since the expansion coefficients $ U^{(2)} $ and $ U^{(3)} $ are related to
elastic constants, as summarized in Table \ref{graphanechair} for the C-graphane and in Table \ref{graphaneboat} for the B- and W-graphane, a straightforward
fit of Eq.(\ref{fit3}) has provided the full set of third-order elastic constants. 

The results have been reported in Table \ref{table:nonlin} where only the values of the independent elastic constants appearing in Eqs. (\ref{completehex}), (\ref{completetrigo}) and (\ref{completeortho}) are reported.
We note that graphene and B-graphane are characterized by an inverted anisotropy: while $ C_{111}<C_{222} $ for graphene, we found  $ C_{222}>C_{111} $ for B-graphane. On the contrary, W-graphane has the same anisotropy of graphene ($ C_{111}<C_{222} $), but a larger $ |C_{111}-C_{222}| $ difference.
So, it is interesting to observe that the different distribution of hydrogen atoms can induce strong qualitative variations for the nonlinear elastic behavior of these structures.

We finally observe that necessarily $C_{444}=0$ for B- and W- graphane because of the orthorhombic symmetry. On the other hand, this nonlinear shear modulus could assume any value for the trigonal lattice. Interesting enough, we have verified that $C_{444}=0$ also for C-graphane. This is due to the additional (with respect to the trigonal symmetry) mirror symmetry of C-graphane.

Similarly to the case of graphene,\cite{lee,cadelano} 
a nonlinear stress-strain relation can be derived for the three graphane conformers:
\begin{equation}
\sigma_{\vec{n}}=E_{\vec{n}}\epsilon_{\vec{n}}+D_{\vec{n}}\epsilon_{\vec{n}}^{2}
\label{av}
\end{equation}
where $E_{\vec{n}}$ and $D_{\vec{n}}$ are, respectively, the Young modulus and an effective nonlinear (third-order) elastic modulus, along the arbitrary direction ${\vec{n}}$, as defined in Sec.~ \ref{sec:linear}.
\begin{figure}[t]
\begin{center}
\includegraphics[width= 0.45\textwidth, angle=0]{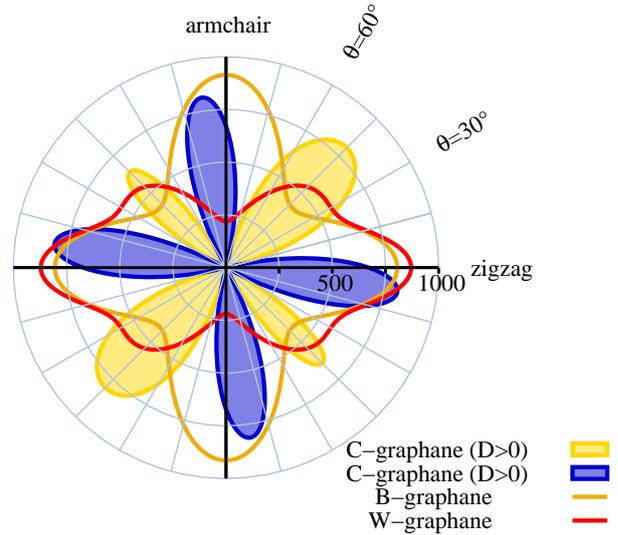}
\caption{\label{nonlinearmoduli} (color online) Polar representation of the nonlinear elastic moduli $ D_{\vec{n}}$ of the three graphane conformers. In the B- and W-graphane cases, $ D_{\vec{n}}\equiv D$ are everywhere negative (softening hyperelesticity), while in the C-graphene one the $ D_{\vec{n}}$ alternates negative and positive values (hardening hyperelesticity). }
\end{center}
\end{figure}
The nonlinear elastic modulus $ D_{\vec{n}}^{(trigo)} $ for the C-graphane (as well as for any trigonal 2D lattice) is given by
\begin{eqnarray}
\nonumber
& & D_{\vec{n}}^{(trigo)}=\tfrac{1}{2}\Bigl[ \nu \left( 1-\nu \right)\left( C_{111}-3C_{112} \right)\,\\\nonumber
& & +\left( 1-\nu \right) \left( 1+\nu^2 \right)C_{111}\\ \nonumber
& & +6cs\left( 1+\nu \right) \left( 1+\nu^2 \right)C_{114}-12cs\left( 1+\nu \right) \nu C_{124}\,\\ \nonumber
& & +3c^2s^2\left( 1-\nu \right) \left( 1+\nu^2 \right)\left( -C_{111}+4C_{144}+C_{112} \right)\,\\ \nonumber
& & +4c^3s^3\left( 1+\nu \right) \left( 1+\nu^2 \right)\left( -3C_{114}+2C_{444}+3C_{124} \right)\,\\ 
& & +8c^3s^3\left( 1+\nu \right)\nu ~\left( -6C_{114}+5C_{444}+6C_{124} \right)\,\Bigr]
\end{eqnarray}
while the corresponding modulus $D_{\vec{n}}^{(ortho)}$ B- and W-graphene is
\begin{eqnarray}
\nonumber
D_{\vec{n}}^{(ortho)}&=&\dfrac{1}{2\Delta^{3}E_{\vec{n}}^{3}} \Bigl[ C_{111}\left( C_{22}c^2 -C_{12}s^2\right)^{3}  \\
\nonumber
& & +C_{222}\left( C_{11}s^2 -C_{12}c^2\right)^{3}\\
\nonumber
& & +3C_{112}\left( C_{11}s^2 -C_{12}c^2\right)\left( C_{22}c^2 -C_{12}s^2\right)^{2}\\
\nonumber
& & +3C_{122}\left( C_{22}c^2 -C_{12}c^2\right)\left( C_{11}s^2 -C_{12}c^2\right)^{2}\\
\nonumber
& & -3C_{166}c^2 s^2\left( C_{22}c^2 -C_{12}s^2\right)\left( \Delta/C_{44}\right)^{2}\\
& & -3C_{266}c^2 s^2\left( C_{11}s^2 -C_{12}c^2\right)\left( \Delta/C_{44}\right)^{2} \Bigr]
\end{eqnarray}
Since $ C_{ijk}<0 $, as shown in Tab.\ref{table:nonlin}, $D_{\vec{n}}^{(ortho)}$ are negative for any direction ( see Fig.~\ref{nonlinearmoduli}), so both B- and W-graphane show an hyperelastic softening behavior. The trigonal C-graphane behaves in a very different way instead. Since the $ C_{114} $ and  $ C_{144} $ are positive, the C-graphane can show an hyperelastic hardening behavior in the angular sectors $5/{12} \pi+ k\pi<\theta <1/{12}+k \pi$ and  $8/{12} \pi+ k\pi<\theta <10/{12}+k \pi$ ($ k\epsilon \mathbb{Z}$).

\section{Conclusions}
In conclusion, present first principles calculations predict that the class of auxetic materials is larger than reported so far, including as well two dimensional hydrocarbons like B-graphane. More precisely, since a negative Poisson ratio is observed for extensions along the zigzag and armchair principal directions, B-graphane is better referred to as an axially auxetic atomic sheet. Moreover, we calculated that the other two conformers, namely the C- and W-graphane, exhibit a vanishingly small value of the Poisson ratio.
The nonlinear elastic behavior of graphane shows peculiar features as well. In particular, we have found that the C-graphane admits both softening and hardening hyperelasticity, depending on the direction of the applied strain.
These features makes graphane a very intriguing material with potentially large technological impact in nanomechanics.

\begin{acknowledgments}
We acknowledge computational support by CYBERSAR (Cagliari, Italy) and CASPUR (Roma, Italy) computing centers. Two of us, P.L.P. and S.G.,  acknowledge financial support by University of Padova (Italy) under project MATHXPRE and by CYBERSAR, respectively.
\end{acknowledgments}


\begin{thebibliography}{99}

\bibitem{sofo} 
J. O. Sofo,  A. S. Chaudhari, and G. D. Barber, Phys. Rev. B {\bf 75}, 153401 (2007).

\bibitem{boukhvalov} 
D. W. Boukhvalov, M. I. Katsnelson, and A. I. Lichtenstein, Phys. Rev. B {\bf 77}, 035427 (2008).

\bibitem{elias} 
D. C. Elias, R.R. Nair, T. M. G. Mohiuddin, S. V. Morozov, P. Blake, M. P. Halsall, A. C. Ferrari, D. W. Boukhvalov, M. I. Katsnelson, A. K. Geim, and K.S. Novoselov, Science {\bf 323}, 610 (2009).

\bibitem{topsakal} 
M. Topsakal, S. Cahangirov, and S. Ciraci, Appl. Phys. Lett. {96}, 091912 (2010).

\bibitem{cadelano} 
E. Cadelano, P. Palla, S. Giordano, and L. Colombo, Phys. Rev. Lett. {\bf 102}, 235502 (2009) and references therein.

\bibitem{evans}
K. E. Evans, M. A. Nkansah, I. J. Hutchinson, S. C. Rogers, Nature \textbf{353}, 124 (1991). 


\bibitem{quantum-espresso}
P. Giannozzi, S. Baroni, N. Bonini, M. Calandra, R. Car, C. Cavazzoni, D. Ceresoli, G. L. Chiarotti, M. Cococcioni, I. Dabo1, A. Dal Corso, S. de Gironcoli, S. Fabris, G. Fratesi, R. Gebauer, U. Gerstmann, C. Gougoussis, A. Kokalj, M. Lazzeri, L. Martin-Samos, N. Marzari, F. Mauri, R. Mazzarello, S. Paolini, A. Pasquarello, L. Paulatto, C. Sbraccia, S. Scandolo, G. Sclauzero, A. P. Seitsonen, A. Smogunov, P. Umari1 and R. M. Wentzcovitch1, J. Phys.: Condens. Matter \textbf{21} 395502 (2009).

\bibitem{vanderbilt}
D. Vanderbilt, Phys. Rev. B \textbf{41}, 7892 (1990).  

\bibitem{kudin}
K. N. Kudin,E. Scuseria and B. I. Yakobson, Phys. Rev. B {\bf 64}, 235406 (2001).

\bibitem{gui}
G. Gui, J. Li, and J. Zhong, Phys. Rev. B \textbf{78}, 075435 (2008).

\bibitem{liu}
F. Liu, P. Ming and J. Li, Phys. Rev. B \textbf{76}, 064120 (2007).

\bibitem{baroni}
S. Baroni, S. de Gironcoli, A. Dal Corso and P. Giannozzi, Rev. Mod. Phys. \textbf{73}, 515 (2001).

\bibitem{washboard} 
V. I. Artyukhov and L. A. Chernozatonskii, J. Phys. Chem. A \textbf{114}, 5389 (2010).


\bibitem{dresselhaus}
R. Saito, G. Dresselhaus, and M. S. Dresselhaus, {\em  Physical Properties of Carbon Nanotubes} (Imperial College
Press, London, 1998).


\bibitem{huntington}
H. B. Huntington, \textit{The elastic constants of crystals} (Academic Press, New York, 1958).

\bibitem{landau}
L. D. Landau and E. M. Lifschitz, {\em Theory of Elasticity} (Butterworth Heinemann, Oxford, 1986).

\bibitem{advmat}
K. E. Evans, A. Alderson, Adv. Mat. {\bf 12}, 617 (2000).

\bibitem{wei}
X. Wei, B. Fragneaud, C. A. Marianetti, and J. W. Kysar, Phys. Rev. B \textbf{80}, 205407 (2009).

\bibitem{flores} 
M. Z. S. Flores, P. A. S. Autreto, S. B. Legoas and D. S. Galvao, Nanotechnology \textbf{20},  465704 (2009).

\bibitem{saito}
R.Saito, G. Dresselhaus and M.S. Dresselhaus, {\em Physical properties of carbon nanotubes} (Imperial College Press, London, 2003). 

\bibitem{lee}
C. Lee, X. Wei, J. W. Kysar, and J. Hone, Science {\bf 321}, 385 (2008).
\end{thebibliography}
\end{document}